\documentclass[acmtog, authorversio]{acmart}
\usepackage{multirow}
\usepackage{graphicx}
\graphicspath{{Figures/}}
\usepackage{subfigure}

\AtBeginDocument{%
  \providecommand\BibTeX{{%
    \normalfont B\kern-0.5em{\scshape i\kern-0.25em b}\kern-0.8em\TeX}}}

\setcopyright{acmcopyright}
\copyrightyear{2025}
\acmYear{2025}
\acmDOI{XXXXXXX.XXXXXXX}

\acmConference[Conference acronym 'XX]{Make sure to enter the correct
  conference title from your rights confirmation emai}{June 03--05,
  2018}{Woodstock, NY}
\acmBooktitle{Woodstock '18: ACM Symposium on Neural Gaze Detection,
 June 03--05, 2018, Woodstock, NY} 
\acmPrice{15.00}
\acmISBN{978-1-4503-XXXX-X/18/06}

\usepackage{hyperref}

\begin{document}

\title{PsyCounAssist: A Full-Cycle AI-Powered Psychological Counseling Assistant System}


\author{Xianghe Liu}
\affiliation{%
  \institution{Beijing PsychTech Technology Co., Ltd.}
  \city{Beijing}
  \country{China}}
\email{xianghe.liu01@gmail.com}

\author{Jiaqi Xu}
\affiliation{%
  \institution{Beijing PsychTech Technology Co., Ltd.}
 \city{Beijing}
 \country{China}
  }
\email{jiaqixu0626@gmail.com}

\author{Tao Sun}
\authornote{Corresponding author.}
\affiliation{%
  \institution{University of Hechi}
 \city{Hechi}
 \country{China}
  }
  \email{2024660033@hcnu.edu.cn}

\renewcommand{\shortauthors}{Liu et al.}

\begin{abstract}

Psychological counseling is a highly personalized and dynamic process that requires therapists to continuously monitor emotional changes, document session insights, and maintain therapeutic continuity. In this paper, we introduce \textbf{PsyCounAssist}, a comprehensive AI-powered counseling assistant system specifically designed to augment psychological counseling practices. PsyCounAssist integrates multimodal emotion recognition combining speech and photoplethysmography (PPG) signals for accurate real-time affective analysis, automated structured session reporting using large language models (LLMs), and personalized AI-generated follow-up support. Deployed on Android-based tablet devices, the system demonstrates practical applicability and flexibility in real-world counseling scenarios. Experimental evaluation confirms the reliability of PPG-based emotional classification and highlights the system's potential for non-intrusive, privacy-aware emotional support. PsyCounAssist represents a novel approach to ethically and effectively integrating AI into psychological counseling workflows.

\end{abstract}

\ccsdesc[500]{Computing methodologies~Artificial intelligence}
\ccsdesc[500]{Applied computing~Life and medical sciences}
\ccsdesc[300]{Security and privacy~Human and societal aspects of security and privacy}

\keywords{Psychological counseling, Multimodal emotion recognition, Photoplethysmography (PPG), Large language models (LLM), AI-assisted counseling, Human-centered AI}




\settopmatter{printacmref=false} 
\renewcommand\footnotetextcopyrightpermission[1]{} 

\maketitle

\section{Introduction}

Psychological counseling is increasingly recognized as a crucial service amid the growing global prevalence of mental health disorders, yet the field continues to face significant resource constraints, notably shortages of trained therapists \cite{chung2023challenges}. To bridge this gap, artificial intelligence (AI) technologies, particularly AI chatbots and automated systems, have been widely explored and implemented to support or even partially replace traditional counseling practices \cite{fitzpatrick2017delivering}\cite{park2023effect}. However, these approaches frequently aim at substituting therapists with conversational agents, which introduces significant ethical concerns, including the lack of empathy, potential biases, and insufficient personalization \cite{zhang2024can}.

Recent research highlights the limitations of existing AI-based counseling systems, specifically their inability to adequately understand nuanced human emotions, respond empathically, and address individual differences in emotional expression \cite{park2023effect}. Moreover, current solutions rarely consider the needs of therapists who require effective, reliable tools to enhance their clinical workflows and reduce administrative burdens, while also safeguarding clients' privacy \cite{chung2023challenges}.

In response to these gaps, we introduce PsyCounAssist, a novel AI-powered full-cycle psychological counseling assistant system designed explicitly to support, rather than replace, therapists and their clients. Our system focuses on three distinct yet interconnected modules aligned with actual counseling scenarios: real-time emotion prediction (REP), automated structured session reporting (ASSR), and personalized follow-up support (PFS).

PsyCounAssist leverages a multimodal fusion approach combining speech and photoplethysmography (PPG) signals, thereby significantly enhancing the accuracy of real-time emotion recognition compared to traditional unimodal methods \cite{ping2024experience}\cite{gyaneshwar2024mental}. Beyond real-time emotional monitoring, PsyCounAssist integrates an innovative session documentation system powered by large language models (LLMs), automating the structured reporting process and substantially decreasing therapists' documentation workloads. Additionally, our system provides a personalized conversational agent that supports clients between counseling sessions. Clients can choose between general or individualized chatbots, the latter leveraging previous session insights to deliver more tailored emotional support and self-help interventions.

Critically, PsyCounAssist emphasizes ethical considerations and privacy protection. Unlike conventional AI counseling systems, our design explicitly allows emotional monitoring without obligatory audio recording or transcription, thus ensuring clients' confidentiality and data security \cite{chung2023challenges}.

In summary, the primary contributions of this work include: (1) proposing PsyCounAssist, a novel system designed explicitly as a therapist-support tool rather than a replacement, addressing key ethical and practical limitations of existing approaches; (2) introducing an advanced multimodal fusion method for emotion recognition that combines speech and PPG signals to enhance accuracy and privacy protection; and (3) demonstrating practical, deployable tools such as automated structured reporting and personalized follow-up support, aimed at significantly improving therapeutic efficiency and client engagement.

\begin{figure*}
    \centering
    \includegraphics[width=2.0\columnwidth]{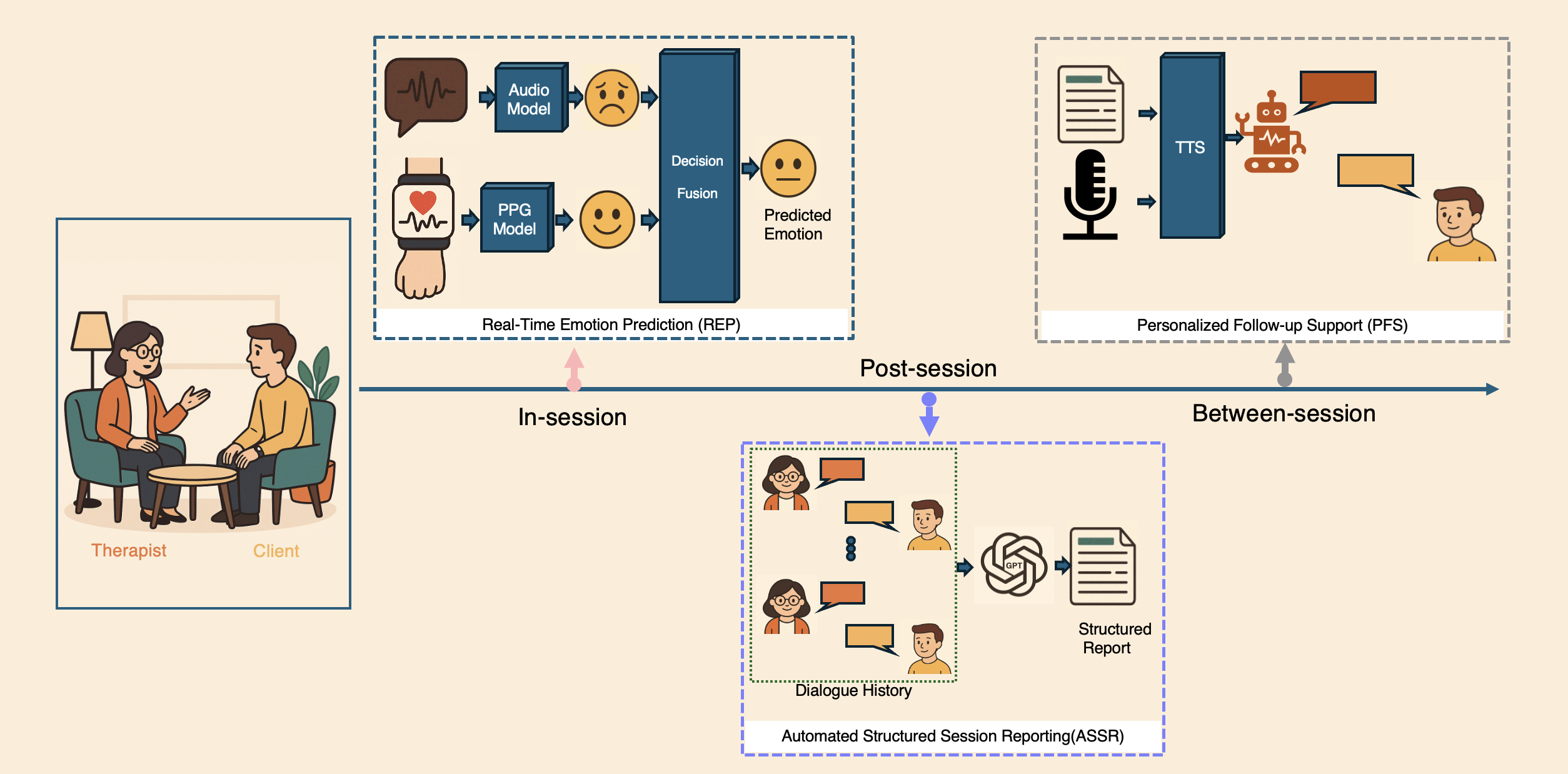}
    \caption{Overview of proposed PsyCounAssist framework.}
    \label{fig:overview}
\end{figure*}














\section{Related Work}

\textbf{AI-Assisted Mental Health Counseling}. In recent years, artificial intelligence (AI) has been successfully applied across various domains, including image recognition and text analysis, progressively merging with psychology to address practical challenges. Within psychological research, AI applications predominantly focus on affective computing, leveraging multi-source data to predict emotional states, valence levels, or mental health indicators. In mental health counseling, notable implementations include chatbot systems and diagnostic recommendation tools designed to simulate certain aspects of human counselor interactions. For instance, PsyDT \cite{PsyDT} employs large-scale corpora emulating professional therapists' conversational styles to train large language models (LLMs), effectively replicating specific counseling dialogues and interaction logic. Similarly, XiaoIce \cite{xiaoIce}, an AI-based companionship system, provides emotional support through everyday conversations tailored to user preferences, accommodating sustained emotional engagement. However, despite significant progress, these systems generally fail to integrate seamlessly into actual counseling practices, neglecting the intrinsic human-to-human empathetic interaction essential to effective psychological counseling. Hence, constructing supportive rather than substitutive AI systems is vital to genuinely empower psychological counseling. Additionally, real-world counseling sessions often involve privacy concerns and expression constraints, leading clients to resist providing complete textual or facial data. Thus, incorporating non-verbal cues as a low-intrusiveness supplement to assist therapists in real-time emotional detection becomes imperative.





\textbf{Multimodal Emotion Recognition in Mental Health}. Multimodal modeling, particularly combining speech and physiological signals, is becoming mainstream for non-invasive, real-time emotion recognition in psychological applications. In speech-based emotion recognition, significant advancements such as the Emotion2Vec model have demonstrated outstanding performance on multimodal datasets like RECOLA \cite{ringeval2013introducing}, achieving state-of-the-art Weighted Accuracy (WA) of 79.4 \%, making it suitable for dynamic assessment and real-time assistance during therapy sessions. Regarding physiological signals, photoplethysmography (PPG) has emerged as a practical, wearable approach extensively used for emotion recognition. PPG-derived features, including heart rate (HR) and heart rate variability (HRV), are strongly correlated with autonomic nervous system activities, effectively indicating fluctuations in anxiety, stress, and other emotional states. These physiological indicators provide reliable data for developing non-invasive multimodal emotion recognition systems. Hence, future psychological counseling AI systems should transition from single-modal simulations toward multimodal fusion approaches, significantly enhancing their capacity to perceive and understand complex human emotional behaviors.





\textbf{LLM-based Automated Counseling Support}. Large language models (LLMs), such as GPT-4, have demonstrated exceptional reasoning and summarization capabilities in language understanding and generation tasks, making them highly suitable for structuring and summarizing counseling session content. Typically, counselors must manually document key conversational insights, a process prone to omission and inefficiency. Automated structured documentation from session recordings substantially reduces counselor workload, enhancing productivity and service quality. Current leading LLMs, including GPT-4, excel in extracting critical information from lengthy counseling sessions (45–60 minutes) and generating organized summaries. Nonetheless, cloud-deployed models pose potential risks regarding user privacy, particularly sensitive within mental health contexts. To address privacy concerns, locally deployable open-source models like DeepSeek present more appropriate alternatives. These models facilitate voice transcription and structured conversation extraction without external servers, simultaneously ensuring privacy and intelligent performance, thus providing practical solutions for counseling support. Additionally, although companion AI systems like XiaoICE are not classified under formal psychological counseling, their emotional support role is significant. Advanced customization utilizing clients' historical counseling summaries and stylistic preferences could create familiar, personalized digital companions, aiding emotional continuity and psychological stability. Technologically, with client authorization, structured counseling records and tailored prompt engineering can generate personalized supportive content. Combined with advanced voice cloning models such as SparkTTS and Bark, these AI systems can emulate specific counselor voices, offering more authentic and trustworthy interactive experiences for clients.






\section{System Overview}
\label{sec:method}


\textcolor{black}{We propose \textbf{PsyCounAssist}, an integrated AI-driven framework for enhancing psychological counseling through real-time affective assessment, automated session documentation, and personalized post-session support. The system consists of three functionally distinct yet interconnected modules: (1) Real-time Emotion Prediction (REP), (2) Automated Structured Session Reporting (ASSR), and (3) Personalized Follow-up Support (PFS). The overall system workflow is illustrated in Figure \ref{fig:overview}.}


\subsection{Real-time Emotion Prediction (REP) }

\subsubsection{Data Collection}
\textcolor{black}{The Real-time Emotion Prediction (REP) module is designed to infer the client’s emotional state during active counseling sessions using multi-modal input. To accommodate varying session contexts and privacy considerations, the system flexibly supports emotion recognition based on \textbf{speech signals}, \textbf{physiological signals} (i.e., photoplethysmography, PPG), or their \textbf{combination}. The specific modality or fusion strategy is determined according to the client’s informed consent and device availability.}\par

\textcolor{black}{Speech data is captured via in-room or pad microphones, while PPG signals are collected through wearable sensors such as wristbands. When both modalities are available and permitted, their outputs are integrated via a decision-level fusion mechanism to improve prediction robustness. This flexible configuration ensures broad applicability while aligning with ethical and privacy-preserving principles.}

\subsubsection{Emotion Prediction}
\paragraph{Speech-based Emotion Recognition}

\textcolor{black}{For the audio modality, we adopt Emotion2Vec\cite{emotion2vec} as the core emotion recognition model. The model projects speech segments into a continuous affective vector space and has demonstrated strong generalization across several public affective speech corpora. In our use case, the model is fine-tuned for three primary emotional classes: sad, neutral, and positive, which are aligned with typical affective states in therapeutic dialogue.}\par
\textcolor{black}{The audio processing pipeline includes:}

\textcolor{black}{\noindent{Voice Activity Detection (VAD)}: applied to segment effective speech intervals from ambient noise and silence;
}

\textcolor{black}{\noindent{KMeans++ Clustering}: used for unsupervised partitioning of speech segments, aiding in speaker differentiation;}

\textcolor{black}{\noindent{Speaker Attribution}: achieved by comparing cluster characteristics with a pre-constructed database of counselor voice samples, allowing extraction of client-only speech for emotion inference.}

\textcolor{black}{This preprocessing ensures the downstream model focuses on client expressions, filtering out therapist utterances and irrelevant segments.}

\paragraph{Physiological Modeling via PPG}

\textcolor{black}{To complement the limitations of speech signals—e.g., suppressed affect expression or deceptive tone—we incorporate photoplethysmography (PPG) as a secondary modality. PPG signals are acquired via non-invasive wrist-worn devices and analyzed in terms of:
Heart Rate (HR),Heart Rate Variability (HRV).}

\textcolor{black}{To build a reliable mapping between physiological responses and emotional states, we design and implement a controlled emotion elicitation experiment. Participants are exposed to audiovisual stimuli curated to induce specific affective states (e.g., sadness, calmness), during which real-time PPG signals were collected. After each elicitation block, participants self-reported their experienced emotions using a valence-arousal grid, which served as ground truth for supervised learning.The collected data were then used to train a set of machine learning models (e.g., SVM, LightGBM) to classify emotional states based on HR and HRV features.
}

\subsubsection{Decision-level Fusion Strategy}

To enhance robustness under signal-specific limitations and real-world uncertainty, we adopt a rule-based decision-level fusion strategy to integrate emotion predictions from the speech and PPG modalities. The final affective state is determined under three conditions: speech-only, PPG-only, and multi-modal availability. Let us denote the key variables and corresponding fusion rules as follows.

\paragraph{Speech-only Condition.}
Let $\hat{y}_s^t = \arg\max_{c \in \mathcal{C}} \, P_s^t(c)$ denote the predicted class at time $t$ from the speech model, where $P_s^t(c)$ is the posterior probability of class $c \in \mathcal{C} = \{\text{sad}, \text{neutral}, \text{positive}\}$. Then:

\begin{itemize}
    \item If $\hat{y}_s^t = \text{neutral}$ and $P_s^t(\text{sad}) > 0$, override label to \textbf{sad}.
    \item Otherwise, use the class with the highest posterior.
\end{itemize}

This rule accounts for low-energy or ambiguous speech that may be misclassified as neutral while still containing affective signals (e.g., sadness).

\paragraph{PPG-only Condition.}
Let $\mu_t$ denote the mean SCR (skin conductance response) at time $t$, computed from GSR features derived from the PPG signal. We define a cumulative affect score $S_p \in \mathbb{R}$, updated heuristically as:

\[
\Delta = \frac{\mu_t - \mu_{t-1}}{\max(\mu_t, \mu_{t-1})}
\]

\[
S_p \leftarrow S_p + \lambda \cdot m \quad \text{if } \Delta > \theta
\]
\[
S_p \leftarrow S_p - \lambda \cdot (1 - m) \quad \text{if } \Delta < -\theta
\]

where $\lambda$ is a fixed update constant (default: 0.5), $\theta$ is the SCR reactivity threshold (default: 0.3), and $m \in [0,1]$ is a confidence coefficient.

The final emotional label is assigned as:
\begin{itemize}
    \item $S_p > \delta_1 \Rightarrow \text{positive}$,
    \item $S_p < -\delta_1 \Rightarrow \text{sad}$,
    \item otherwise $\text{neutral}$,
\end{itemize}
with $\delta_1 = 1.0$ by default.

\paragraph{Multi-modal Condition.}
When both modalities are available, a weighted average is used to compute the final emotional distribution:

\[
P_f^t(c) = \alpha \cdot P_s^t(c) + (1 - \alpha) \cdot P_p^t(c), \quad \forall c \in \mathcal{C}
\]

Here, $P_s^t(c)$ and $P_p^t(c)$ are the class probabilities from speech and PPG models respectively, and $\alpha \in [0,1]$ is the fusion weight. In default settings:
\begin{itemize}
    \item $\alpha = 0.7$ when speech quality is high,
    \item $\alpha = 0.3$ when speech is noisy or unavailable.
\end{itemize}

The final label is computed as $\hat{y}_f^t = \arg\max_c \, P_f^t(c)$.


\subsubsection{Output Interface and Alert Mechanism}

To support therapist awareness and real-time adaptation during counseling sessions, the REP module outputs emotional predictions via an integrated visualization interface in the counselor’s application (e.g., tablet or workstation). Affective states are updated at fixed intervals (e.g., every 60 seconds) and presented in the following manner:

\begin{itemize}
    \item \textbf{Color-coded status bar:} representing discrete emotional states, such as blue (sad), green (neutral), and yellow (positive);
    \item \textbf{Trend indicators:} displaying affective changes over time (e.g., upward/downward arrows);
    \item \textbf{Popup alerts:} triggered under two specific conditions:
    \begin{enumerate}
        \item sustained low-valence states (e.g., continuous sadness predictions), or
        \item abrupt emotional shifts (e.g., valence change exceeding a predefined threshold).
    \end{enumerate}
\end{itemize}

The interface is designed to be minimally intrusive and context-aware, ensuring that the counselor's focus remains primarily on the client, while still benefiting from dynamic affective feedback.

\subsubsection{Module Innovations and Ethical Considerations}

The REP module introduces multiple design innovations and adheres to ethical principles relevant to real-world psychological counseling:

\begin{itemize}
    \item \textbf{Flexible modality selection:} The system supports unimodal or multimodal emotion inference depending on data availability and user consent, enabling personalization per client or institution.
    
    \item \textbf{Decision-level fusion:} Instead of feature-level fusion, we adopt a rule-based strategy that enhances interpretability and robustness, particularly under missing or low-quality signals.
    
    \item \textbf{Low-latency prediction:} All computations are optimized for real-time inference with latency under 1 second, allowing in-session feedback without disrupting counselor flow.
    
    \item \textbf{Privacy-preserving deployment:} All raw signal processing is performed locally or within secure institutional infrastructure. Neither speech nor biometric data is transmitted externally.
    
    \item \textbf{Human-in-the-loop design:} The system provides emotion predictions as \textit{decision support} rather than definitive labels. Final interpretations and actions remain entirely under the therapist’s discretion.
\end{itemize}

These features ensure that the REP module remains clinically relevant, ethically sound, and adaptable to diverse therapeutic contexts.

\subsection{Automated Structured Session Reporting (ASSR)}
The \textbf{Automated Structured Session Reporting (ASSR)} module is responsible for summarizing and structuring the contents of each counseling session using large language models (LLMs). It aims to reduce the post-session documentation burden for counselors and enhance the traceability and interpretability of psychological processes.

\subsubsection{Input and Model Architecture}

Upon completion of a counseling session, the audio content is either transcribed via an automatic speech recognition (ASR) system or annotated manually. The resulting textual transcript serves as input to a few-shot prompted LLM (e.g., GPT-4\cite{gpt4}, DeepSeek\cite{deepseekai2024deepseekv3technicalreport},), which is fine-tuned through prompt engineering to generate structured session reports. Key functions include:

\begin{itemize}
    \item \textbf{Summarizing} the overall dialogue content;
    \item \textbf{Extracting} emotionally salient segments and critical discussion points;
    \item \textbf{Highlighting} psychological markers such as emotional regulation, relationship dynamics, and cognitive shifts.
\end{itemize}

The model utilizes template-driven prompts with role context (e.g., ``counselor'', ``client'') and time-tagged content to ensure factual consistency and structured output.

\subsubsection{Output Format}

The generated report consists of standardized sections designed to capture both clinical observations and client narratives. These typically include:

\begin{itemize}
    \item \textbf{Session Context:} client background and presenting issues;
    \item \textbf{Exploration Highlights:} key dialogue threads, emotional responses, and therapeutic reflections;
    \item \textbf{Observed Progress:} behavioral, emotional, and cognitive shifts since previous sessions;
    \item \textbf{Follow-up Suggestions:} therapist recommendations for individual or systemic intervention;
    \item \textbf{Summary:} synthesis of therapeutic status and engagement outlook.
\end{itemize}





\subsection{Personalized Follow-up Support (PFS)}

The \textbf{Personalized Follow-up Support (PFS)} module enables post-session engagement through AI-generated, context-aware follow-up messages. These messages are designed to provide continued emotional support and therapeutic continuity outside of scheduled sessions, without requiring real-time counselor input.

\subsubsection{Dialogue Framework and Context Conditioning}

PFS utilizes a large language model (LLM)-driven response generation framework tailored for mental health scenarios. To ensure contextual relevance and therapeutic consistency, each response is conditioned on:

\begin{itemize}
    \item \textbf{Structured session summaries} produced by the ASSR module;
    \item \textbf{Recent emotional trends} inferred from REP output;
    \item \textbf{Client-specific therapeutic goals} (e.g., emotional regulation, cognitive reframing).
\end{itemize}

Responses are generated using customized prompt templates that align with common therapeutic paradigms such as cognitive behavioral therapy (CBT) or supportive counseling.

\subsubsection{Natural AI Voice and Delivery Channel}

The generated messages are converted to audio using a generic \textbf{AI voice synthesis engine}, which produces neutral and emotionally consistent spoken messages. This approach avoids direct replication of counselor identity while preserving the benefits of natural-sounding speech. Clients receive these voice messages via a secure mobile or web interface, supporting:

\begin{itemize}
    \item \textbf{Asynchronous delivery}, allowing users to engage at their own pace;
    \item \textbf{Audio and textual formats}, enabling flexible consumption;
    \item \textbf{Minimal interaction burden}, with passive reception of supportive prompts.
\end{itemize}

\subsubsection{Application Scenarios}

PFS supports various use cases, including daily check-ins, reminders for emotion regulation techniques, and motivational messages. A sample output may resemble:

\begin{quote}
\textit{``Hi, just checking in with you today. Remember to take a few minutes to breathe and slow down if you're feeling overwhelmed. You're doing your best—and that matters.''}
\end{quote}

All outputs are based on predefined ethical triggers and client-approved conditions. Messages are short, emotionally supportive, and aligned with the tone of previous sessions.

\subsubsection{System Properties and Ethical Design}

\begin{itemize}
    \item \textbf{Low-intrusion, high-relevance:} Offers continued presence without replicating human counselors;
    \item \textbf{Therapeutically aligned:} Integrates prior interaction history and session intent into each response;
    \item \textbf{Privacy-first design:} All processing occurs locally or within secured clinical infrastructure;
    \item \textbf{Non-identifiable AI voice:} Prevents anthropomorphizing or confusion about human involvement;
    \item \textbf{Scalable and configurable:} Frequency, tone, and content type can be adjusted per user preferences.
\end{itemize}

Through these features, the PFS module extends therapeutic engagement into everyday contexts, reinforcing psychological safety and continuity while respecting ethical and emotional boundaries.

\section{Pilot Experiment}
\label{sec:evaluation}

\subsection{Dataset and Setup}

To evaluate the feasibility of emotion recognition using physiological signals alone, we conducted a binary classification experiment leveraging photoplethysmography (PPG) data. The dataset was collected from 30 participants (22 male, 8 female; average age: 33 ± 1.25 years), each equipped with a wrist-worn wearable device sampling at 100 Hz.

Participants were exposed to audiovisual stimuli designed to elicit distinct emotional states, specifically “\textit{sad}” and “\textit{relax}” conditions. After each stimulus block, participants self-reported their emotional state using a valence-arousal grid, which served as ground-truth annotation. The dataset was then labeled accordingly and split into training (70\%), validation (20\%), and testing (10\%) sets.


\subsection{Model and Feature Engineering}

We extracted key physiological features from the raw PPG signals, focusing on heart rate (HR) and heart rate variability (HRV) metrics known to correlate with autonomic nervous system activity and affective states. A variety of traditional machine learning classifiers were evaluated, including Random Forest, Gradient Boosting, AdaBoost, Support Vector Machine, and Naive Bayes.

\subsection{Results}

The top-performing models on validation and test sets are summarized in Table~\ref{tab:ppg_model_results}. Notably, the Random Forest classifier achieved the highest performance, with \textit{F1 scores of 0.957 and 0.964} on the validation and test sets, respectively. This indicates a strong discriminative power of PPG signals in distinguishing between low-valence and baseline affective states.


\begin{table}[ht]
\centering
\caption{Performance of top 5 models on validation and test sets (PPG only, labels: \textit{sad} vs \textit{relax})}
\begin{tabular}{lcccc}
\toprule
\textbf{Model} & \textbf{Dataset} & \textbf{Accuracy} & \textbf{F1 Score} \\
\midrule
Random Forest & Validation & 0.954 & 0.957 \\
              & Test       & 0.963 & 0.964 \\
Gradient Boosting & Validation & 0.844 & 0.858 \\
                  & Test       & 0.847 & 0.858 \\
AdaBoost & Validation & 0.711 & 0.747 \\
         & Test       & 0.717 & 0.749 \\
Support Vector Machine & Validation & 0.559 & 0.708 \\
                       & Test       & 0.553 & 0.701 \\
Naive Bayes & Validation & 0.558 & 0.707 \\
            & Test       & 0.554 & 0.702 \\
\bottomrule
\end{tabular}
\label{tab:ppg_model_results}
\end{table}

These results demonstrate the \textit{viability of PPG-based emotion recognition}, particularly in scenarios where speech signals are unavailable, restricted by privacy concerns, or degraded by noise. The superior performance of tree-based models such as Random Forest suggests that non-linear relationships between physiological features and emotional states are effectively captured through ensemble learning.



\section{Discussion}

The development and deployment of PsyCounAssist highlight the growing potential of AI technologies to augment psychological counseling, rather than replace human therapists. Our system embraces a therapist-centric design philosophy, prioritizing decision support, interpretability, and ethical integrity.

One of the key strengths of PsyCounAssist lies in its modular, privacy-preserving architecture. By supporting flexible modality configurations—speech-only, PPG-only, or multimodal—the system can adapt to various client contexts, privacy constraints, and hardware limitations. This flexibility is particularly valuable in real-world scenarios where environmental noise, cultural hesitancy toward verbal expression, or data sensitivity may limit traditional approaches.

Furthermore, our adoption of decision-level fusion enhances robustness and transparency. Unlike feature-level fusion, which often lacks interpretability, decision-level fusion allows therapists to better understand and validate the system’s outputs. The inclusion of a human-in-the-loop framework ensures that emotional inferences are always presented as assistive suggestions, not automated diagnoses, preserving clinical judgment and ethical responsibility.

However, several challenges remain:

\begin{itemize}
    \item \textbf{Generalizability:} Emotion recognition models may not perform consistently across diverse cultural or linguistic populations. Variability in emotional expression norms can reduce cross-domain robustness, requiring future adaptation using diverse training data;
    \item \textbf{Follow-up interactivity limitations:} Although the Personalized Follow-up Support (PFS) module offers tailored audio messages, its current one-way delivery limits real-time responsiveness, especially in urgent or acute scenarios;
    \item \textbf{Clinical validation gap:} The current system has yet to be tested in formal clinical settings, especially among vulnerable populations such as adolescents, individuals with severe affective disorders, or clients with limited digital literacy.
\end{itemize}

Addressing these limitations is essential for broader clinical adoption and long-term reliability.

\section{Future Work}

To further enhance PsyCounAssist's effectiveness and applicability, several directions are proposed for future development:

\subsection{Short-Term Goals}
\begin{itemize}
    \item \textbf{Text-based Emotion Recognition:} Incorporate real-time emotion inference from text (e.g., journaling, chat messages) to complement audio/physiological modalities, especially in low-resource environments.
    \item \textbf{Explainability Enhancements:} Integrate interpretable visualizations (e.g., attention heatmaps, rationale generation) for emotion predictions and LLM outputs to increase therapist trust and system transparency.
    \item \textbf{Real-time Risk Detection:} Design dynamic crisis detection protocols to identify high-risk affective states (e.g., signs of acute distress), with escalation options such as supervisor alerts or emergency contact prompts.
\end{itemize}

\subsection{Long-Term Goals}
\begin{itemize}
    \item \textbf{Cross-cultural and Multilingual Adaptation:} Train emotion and language models on geographically and culturally diverse datasets to ensure inclusiveness and global usability.
    \item \textbf{Longitudinal Progress Tracking:} Leverage multi-session data to monitor therapy progress over time and support predictive modeling of therapeutic outcomes, such as relapse risk or emotional stagnation.
    \item \textbf{EHR Integration:} Enable seamless export of structured session summaries to electronic health record (EHR) platforms, enhancing interoperability and clinical documentation efficiency.
\end{itemize}

Through these enhancements, PsyCounAssist aims to evolve from a demo-stage prototype into a clinically validated, ethically grounded, and globally scalable solution for intelligent mental health care.

\section{Conclusion}
\label{sec:conclusion}

We present PsyCounAssist, a full-cycle, AI-powered assistant system designed to support psychological counseling through real-time emotion tracking, automated session documentation, and personalized follow-up communication. Unlike conventional AI systems that attempt to replace human therapists, PsyCounAssist adopts a therapist-centric approach—augmenting professional judgment with ethically designed, privacy-aware tools.

Our system integrates multimodal emotion recognition using speech and PPG signals, structured reporting powered by large language models (LLMs), and AI-generated post-session messages tailored to each client's emotional history. Deployed on portable Android tablets, the system demonstrates practical applicability in real-world counseling scenarios.

The proposed framework offers a scalable and ethically aligned solution for enhancing therapy quality, therapist productivity, and client engagement. Future iterations aim to expand PsyCounAssist’s capabilities to include text-based emotion sensing, cross-cultural adaptation, and longitudinal outcome tracking—bringing us closer to a new paradigm of intelligent, accessible, and trustworthy mental health care.


\clearpage

\bibliographystyle{ACM-Reference-Format}

\bibliographystyle{IEEEtran}
\bibliography{reference}

\end{document}